\documentclass[onecolumn]{IEEEtran}

\ifCLASSINFOpdf
 \usepackage[pdftex]{graphicx}
\else
  \usepackage[dvips]{graphicx}
\fi

\usepackage{graphicx}
\usepackage{epstopdf}
\usepackage{booktabs}
\usepackage{rotating}
\usepackage{listings}
\usepackage[cmex10]{amsmath}
\usepackage{amsfonts}
\usepackage{amssymb}
\usepackage{amsthm}
\usepackage{algorithm}
\usepackage{algorithmicx}
\usepackage{algpseudocode}
\usepackage{morefloats}
\usepackage{cite}
\usepackage{bm}
\usepackage{setspace}
\usepackage[caption=false,font=normalsize,labelfont=sf,textfont=sf]{subfig}
\begin{document}
\bibliographystyle{ieeetran}
\bibliographystyle{ieeetran}
\title{Group Sparse Precoding for Cloud-RAN with Multiple User Antennas}
\author{Zhiyang Liu, Yingxin Zhao, Hong Wu and Shuxue Ding
\thanks{This paper has been published on \emph{Entropy}. doi: 10.3390/e20020144}
\thanks{This work is supported by the National Natural Science Foundation of China under grant nos. 61501262, 61571244 and 61671254, by Tianjin Research Program of Application Foundation and Advanced Technology under grant nos. 15JCYBJC51600 and 16YFZCSF00540.}
\thanks{Correspondence: Y. Zhao, zhaoyx@nankai.edu.cn}
\thanks{Z. Liu, Y. Zhao, H. Wu and S. Ding are with  Tianjin Key Laboratory of Optoelectronic Sensor and Sensing Network Technology, Department of Communication Engineering, Nankai University, 38 Tongyan Road, Jinnan District, Tianjin 300350, China. }}

\maketitle
\begin{abstract}
  Cloud radio access network (C-RAN) has become a promising network architecture to support the massive data traffic in the next generation cellular networks. In a C-RAN, a massive number of low-cost remote antenna ports (RAPs) are connected to a single baseband unit (BBU) pool via high-speed low-latency fronthaul links, which enables efficient resource allocation and interference management. As the RAPs are geographically distributed, group sparse beamforming schemes attract extensive studies, where a subset of RAPs is assigned to be active and a high spectral efficiency can be achieved. However, most studies {assume} that each user is equipped with a single antenna. How to design the group sparse precoder for the multiple antenna users remains little understood, as it requires the joint optimization of the mutual coupling transmit and receive beamformers. This~paper formulates an optimal joint RAP selection and precoding design problem in a C-RAN with multiple antennas at each user. Specifically, we assume a fixed transmit power constraint for each RAP, and investigate the optimal tradeoff between the sum rate and the number of active RAPs. Motivated~by the compressive sensing theory, this paper formulates the group sparse precoding problem by inducing the $\ell_0$-norm as a penalty and then uses the reweighted $\ell_1$ heuristic to find a solution. By adopting the idea of block diagonalization precoding, the problem can be formulated as a convex optimization, and an efficient algorithm is proposed based on its Lagrangian dual. Simulation results verify that our proposed algorithm can achieve almost the same sum rate as that obtained from an exhaustive search.
\end{abstract}

\section{Introduction}

Cloud radio access network (C-RAN) \cite{CRAN2011} is a promising and flexible architecture to accommodate the exponential growth of mobile data traffic in the next-generation cellular network. In a C-RAN, all~the base-band signal processing is shifted to a single base-band unit (BBU) pool \cite{Shi2015}. The conventional base-stations (BSs), however, are replaced by geographically distributed remote antenna ports (RAPs) with only antenna elements and power amplifiers, which are connected to the BBU pool via high-speed low-latency fronthaul links by fiber. Thanks to its simple structure, it is promising to deploy ultra-dense RAPs in a C-RAN with low cost.

With highly dense geographically distributed RAPs, significant rate gains can be expected over that with the same amount of co-located antennas in both the single-user and multi-user cases \cite{Liu2014,Liu2017,Wang2015}. Due to the huge differences among the distance between the user and the geographically distributed RAPs, it has been shown in \cite{Liu2014} that the capacity in the single-user case is crucially determined by the access distance from the user to its closest RAP. This motivates us to investigate whether it is possible to achieve a significant proportion of the sum rate by using a subset of the RAPs. {When the number of active RAPs is small, the other RAPs will operate in the sleeping mode with very low power consumption, which reduces the network power consumption, and significantly improves the energy efficiency \cite{Dai2016, Nguyen2017}. Moreover, it is also able to improve the rate performance under limited fronthaul link capacity, as each RAP will only serve users with small access distances \cite{Zhao2013, Dai2013, Ha2016, Wang2017}.} In the single-user case, it is straightforward to avoid distant RAPs transmitting, as they have little contribution to improving the capacity. In the multi-user case, the problem becomes challenging, as the beamformers of all users should be jointly designed.

To tackle this problem, a branch of sparse beamforming technologies are therefore proposed, where the beamforming vectors are designed to be sparse %We remove the italic here, please confirm.
 with respect to the total number of transmit antennas \cite{Zhao2013,Mehanna2013,Dai2013}. {Intuitively, a sparse beamforming vector implies that the number of active BS antennas is much smaller than the total number of BS antennas, leading to a significant reduction on both active fronthaul links and the circuit power consumption. Typically, the sparse constraint is introduced by imposing the $\ell_0$ norm of the beamforming vector as a regularization of the original objective function. As the $\ell_0$ norm is neither convex nor continuous, the sparse beamforming problem is in general a mixed-integer optimization problem, which is difficult to be globally optimized.} Motivated by the recent theoretical breakthroughs in compressive sensing \cite{Zhang2015}, the sparse beamforming problem is formulated by including the $\ell_1$ norm of the beamforming vectors as a regularization such that the problem becomes convex. By iteratively updating the weights of the $\ell_1$ norm, the sparse beamformer that minimizes the total transmit power can be obtained by iteratively solving a second-order-conic-programming (SOCP) \cite{Zhao2013} or a semi-definite-programming (SDP) \cite{Mehanna2013}. The problem can be further simplified to {an} uplink beamformer design problem via uplink-downlink duality \cite{Dai2013}.

Nevertheless, when each RAP includes multiple antennas, one RAP will be switched off only when all the coefficients in its beamformer are set to be zero. In other words, all antennas at a RAP should be selected or ignored simultaneously, {otherwise the number of active links from the BBU pool to the RAPs and the circuit power consumption cannot be reduced.} {Recently, the group sparse beamforming problem has been proposed where the antennas at the same RAP are restricted to be switched on or off simultaneously \cite{Dai2014a,Shi2014,Shi2015a, Dai2016, Ha2016, Luong2017,Nguyen2017}, which further complicates the problem. Luong et al.  \cite{Luong2017} % Reference citation is not allowed at the begining of the sentence,  we reformatted, please confirm all the same situation.
 {formulated} the sum rate and power consumption tradeoff problem as a mixed-integer-second-order-conic-programming (MI-SOCP) problem to obtain the global optimum by using branch-and-reduce-and-bound (BRB) algorithm, which imposes prohibitively computational complexity when the numbers of the RAPs and the users are large. To reduce the complexity, convex approximations are usually used to make it convex, continuous, and differentiable. For instance,  Dai~and~Yu \cite{Dai2014a} {introduces} a reweighted $\ell_1$ norm of the vector that identifies the transmission power at each RAP to approximate the number of active RAPs, and the non-convex weighted sum rate maximization problem is approximated by weighted minimum mean square error (WMMSE) minimization and can be solved via a quadratical-constrained-quadratic-programming (QCQP). A more commonly used method is to replace the rate constraint by its equivalent signal-to-interference-plus-noise ratio (SINR) constraint, so as to solve the problem via SOCP \cite{Shi2014, Shi2015a, Dai2016, Luong2017}.} Compared to the individual sparse beamforming, the group sparse beamforming further reduces the network power consumption, and the energy efficiency can be improved as well.

So far, most algorithms focus on the situation where each user has a single antenna \cite{Zhao2013,Mehanna2013,Dai2013,Dai2014a,Shi2014,Shi2015a, Dai2016, Luong2017}. As suggested by the multiple-input-multiple-output (MIMO) theory, the capacity increases linearly with the minimum number of transmit and receive antennas \cite{Telatar1999}. In a C-RAN, it is desirable to employ multiple antennas at each user to exploit the potential multiplexing gains, which, however, further complicates the sparse precoder design. {In fact, most techniques \cite{Zhao2013,Mehanna2013,Dai2013,Shi2014,Shi2015a, Dai2016, Luong2017} developed for single-antenna users cannot be directly applied to the multiple-antenna user case. The difficulty originates from the fact that the rate is determined by not only the SINR but also by the power allocated to the multiple sub-channels, and thus the problem cannot be transformed into an SOCP problem as in the group sparse beamformer design. Therefore,  Pan et al. \cite{Pan2017} {proposed} to adopt the reweighted $\ell_1$ norm to make the sparse constraint smooth and use the WMMSE method to make the rate expression convex, and a low-complexity algorithm is proposed to solve the network power consumption minimization problem by exploiting the special structure of the WMMSE approximation. This motivates us to extend some well-structured precoding scheme to C-RAN, and design a group sparse precoding approach with low computational complexity. In particular, we focus on designing a group sparse precoder based on a orthogonal precoding scheme, block diagonalization (BD) \cite{Spencer2004}, which has gained widespread popularity thanks to its low complexity and near-capacity performance when the number of transmit antennas is large \cite{Shen2006,Shen2007,Shim2008,Ravindran2008}. With BD, the receiving beamformer can be directly calculated from the channel gain matrix and the transmit precoder, and the design of the receive beamformer, which is mutually coupled with the transmit beamformer and difficult to be optimized in the multiple antenna user case \cite{Cai2011}, can be further simplified}.

In this paper, we address the joint problem of RAP selection and joint precoder design in a C-RAN with multiple antennas at each user and each RAP. Whereas the problem is typically non-deterministic polynomial-time hard (NP-hard), %Please confirm.
 we show that the problem becomes convex by inducing the reweighted $\ell_1$ norm of a vector that indicates the transmit power at each RAP as a regularization. Based on its Lagrangian dual problem, we propose an algorithm by iteratively updating the weights of the $\ell_1$ norm to generate a sparse solution. Simulation results verify that the proposed algorithm can achieve almost the same sum rate as {that from exhaustive search}.%meaning retained?

The rest of this paper is organized as follows: Section 2 introduces the system model and formulates the problem. Section 3 proposes an iterative algorithm to solve the group sparse precoding problem. {The complexity analysis of the proposed algorithm is presented in Section 4.} Simulation results are presented and discussed in Section 5. Section 6 concludes this paper.

{\textit{Notation:} Italic letters denote scalars, and boldface upper-case and lower-case letters denote matrices and vectors, respectively. $\|\mathbf{x}\|_p$ denotes the $\ell_p$ norm of vector $\mathbf{x}$. $\mathbf{X}^T$, $\mathbf{X}^{\dag}$, $\textmd{Tr}\{\mathbf{X}\}$ and $\det\{\mathbf{X}\}$ denote the transpose, conjugate transpose, trace and determinant of matrix $\mathbf{X}$, respectively. $\textrm{diag}(a_1,\dots,a_N)$ denotes an $N\times N$ diagonal matrix with diagonal entries $\{a_i\}$. $\mathbf{I}_N$ denotes an $N\times N$ identity matrix. $\mathbf{0}_{N\times M}$ and $\mathbf{1}_{N\times M}$ denote $N\times M$ matrices with all entries zero and one, respectively. $|\mathcal{X}|$~denotes the cardinality of set $\mathcal{X}$. $\lceil\cdot\rceil$ and $\mathbb{E}[\cdot]$ denote the ceiling and expectation operators,~respectively.}

\section{System Model and Problem Formulation}\label{s2}

\begin{figure}[H]
\begin{center}
\includegraphics[width=.4\textwidth]{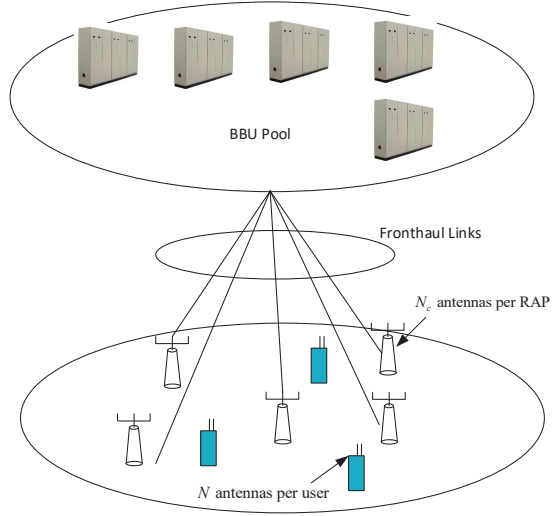}
\caption{Graphic illustration of a Cloud radio access network {(C-RAN)}; BBU: baseband unit; RAP: remote antenna port. }\label{Fig cran} %Please confirm.
\end{center}
\end{figure}

Consider a C-RAN with a set of remote antenna ports (RAPs), denoted as $\mathcal{L}$, and a set of users, denoted as $\mathcal{K}$, with $|\mathcal{L}|=L$ and $|\mathcal{K}|=K$, as shown in Figure~\ref{Fig cran}. Suppose that each RAP is equipped with $N_c$ antennas, and each user is equipped with $N$ antennas. {Then we have a total number of $M=N_c L$ BS antennas.} The baseband units (BBUs) are moved to a single BBU pool which are connected to the RAPs via high-speed fronthaul links, such that the BBU pool has access to the perfect channel state information (CSI) between the RAPs and the users, and the signals of all RAPs can be jointly processed. With a high density of geographically distributed RAPs, the access distances from each user to the RAPs varies significantly, and the distant RAPs have little contribution to improve the capacity. This motivates us to find a subset of RAPs that can provide near optimal sum rate~ performance.

In particular, let $\mathcal{A}\subseteq \mathcal{L}$ denote the set of active RAPs, with $|\mathcal{A}|=A$. To utilize the multiplexing gains from the use of multiple user antennas, we assume that $AN_c\geq N$. The received signal at user $k$ can be then modeled as
\begin{equation}\label{System Model}
\mathbf{y}_k=\mathbf{H}_{k,\mathcal{A}}\mathbf{x}_{k,\mathcal{A}}+\sum_{j\ne k}\mathbf{H}_{k,\mathcal{A}}\mathbf{x}_{k,\mathcal{A}}+\mathbf{z}_k,
\end{equation}
where $\mathbf{x}_{k,\mathcal{A}}\in\mathbb{C}^{AN_c\times 1}$  and $\mathbf{y}_k\in\mathbb{C}^{N\times 1}$ denote the transmit and receive signal vectors, respectively. $\mathbf{H}_{k,\mathcal{A}}\in\mathbb{C}^{N\times AN_c}$ is the channel gain matrix between the active RAPs and user $k$. $\mathbf{z}_k$ denotes the additive noise, which is modeled as a Gaussian random vector with zero mean and covariance $\sigma^2 \mathbf{I}$. With linear precoding, the transmit signal vector $\mathbf{x}_{k,\mathcal{A}}$ can be expressed as
\begin{equation}
\mathbf{x}_{k,\mathcal{A}}=\mathbf{T}_{k,\mathcal{A}} \mathbf{s}_k,
\end{equation}
where $\mathbf{T}_{k,\mathcal{A}}\in\mathbb{C}^{AN_c\times N}$ is the precoding matrix. $\mathbf{s}_k\in\mathbb{C}^{N\times 1}$ is the information bearing symbols. It is assumed that Gaussian codebook is used for each user at the transmitter, and therefore $\mathbf{s}_k\sim\mathcal{CN}\left(\mathbf{0},\mathbf{I}\right)$. The transmit covariance matrix for user $k$ can be then written as $\mathbf{S}_{k,\mathcal{A}}=\mathbb{E}\left[\mathbf{x}_{k,\mathcal{A}}\mathbf{x}_{k,\mathcal{A}}^\dag\right]$. It is easy to verify that $\mathbf{S}_{k,\mathcal{A}}=\mathbf{T}_{k,\mathcal{A}}\mathbf{T}_{k,\mathcal{A}}^\dag$. The sum rate can be written from (\ref{System Model}) as
\begin{equation}\label{sum rate}
R=\sum_{k\in\mathcal{K}} \log_2 \det\left(\mathbf{I}_N+\frac{\mathbf{H}_{k,\mathcal{A}}\mathbf{S}_{k,\mathcal{A}}\mathbf{H}_{k,\mathcal{A}}^\dag}{\sigma^2\mathbf{I}_N+\sum_{j\ne k}\mathbf{H}_{k,\mathcal{A}}\mathbf{S}_{j,\mathcal{A}}\mathbf{H}_{k,\mathcal{A}}^\dag}\right).
\end{equation}

Note that it is difficult to find the optimal linear precoder that maximizes the sum rate $R$ due to the non-convexity of (\ref{sum rate}), and the mutual coupling of the transmit and receive beamformers makes it difficult to jointly optimize the beamformers \cite{Cai2011}. In this paper, we assume that block diagonalization (BD) is adopted, where {the desired signal is projected to the null space of the channel gain matrices of all the other users,} such that $\mathbf{H}_{k,\mathcal{A}}\mathbf{x}_{j,\mathcal{A}}=\mathbf{0}$, or equivalently $\mathbf{H}_{k,\mathcal{A}}\mathbf{S}_{j,\mathcal{A}}\mathbf{H}_{k,\mathcal{A}}^\dag=\mathbf{0}$ for all $j\ne k$. With BD, the sum rate can be obtained as
\begin{equation}
R=\sum_{k\in\mathcal{K}} \log_2 \det\left(\mathbf{I}_N+\frac{1}{\sigma^2}\mathbf{H}_{k,\mathcal{A}}\mathbf{S}_{k,\mathcal{A}}\mathbf{H}_{k,\mathcal{A}}^\dag\right).
\end{equation}

As the signals come from more than one RAP, they need to satisfy a set of per-RAP power constraints, i.e.,
\begin{equation}
\text{Tr}\left\{\mathbf{B}_l\sum_k\mathbf{S}_{k,\mathcal{A}}\right\}\leq P_{l,\max},\forall l\in\mathcal{A},
\end{equation}
where $P_{l,\max}$ denotes the per-RAP power constraint. $\mathbf{B}_l=\text{diag}\{\mathbf{b}_l\}$ is a diagonal matrix, whose diagonal entries is defined as
\begin{equation}
\mathbf{b}_l=[\underbrace{0,\cdots,0}_{N_c(l-1)},\underbrace{1,\cdots,1}_{N_c},\underbrace{0,\cdots,0}_{M-N_c l}].
\end{equation}

This paper focuses on the tradeoff between the sum rate and the group sparsity. In particular, the problem can be formulated as the following optimization problem: % subequations are not allowed according to the journal style, please confirm the equations (7)-(8) and all the citations.

%\begin{subequations}\label{Primal}
\begin{align}
&\mathop{\text{maximize}}\limits_{\mathcal{A},\{\mathbf{S}_{k,\mathcal{A}}\}}&& \sum_{k\in\mathcal{K}}\log_2\det\left(\mathbf{I}_N+\frac{1}{\sigma^2}\mathbf{H}_{k,\mathcal{A}}\mathbf{S}_{k,\mathcal{A}}\mathbf{H}_{k,\mathcal{A}}^{\dag}\right) -\eta|\mathcal{A}| &&\label{Primal Constraint 1}\\
&\text{s.t.} && \mathbf{H}_{j,\mathcal{A}}\mathbf{S}_{k,\mathcal{A}}\mathbf{H}_{j,\mathcal{A}}=\mathbf{0}&& \forall j,k\in\mathcal{K},j\ne k \label{Primal Constraint 2}\\
&&& \text{Tr}\left\{\mathbf{B}_l\sum_k\mathbf{S}_{k,\mathcal{A}}\right\}\leq P_{l,\max} && \forall l\in\mathcal{A}\label{Primal Constraint 3} \\
&&& \mathbf{S}_{k,\mathcal{A}}\succeq 0&& \forall k\in\mathcal{K}\label{Primal Constraint 4}
\end{align}
%\end{subequations}
where (\ref{Primal Constraint 2}) is the zero-forcing (ZF) constraint, which ensures that the inter-user interference can be completely eliminated at the optimum. $\eta\geq 0$ is the tradeoff constant, which controls the sparsity of the solution, and thus the number of active RAPs. With $\eta=0$, the problem reduces to a BD precoder optimization problem with per-RAP power constraint. The group sparsity can be improved by assigning a larger $\eta$.

Note that the optimization problem {defined in (\ref{Primal Constraint 1})--(\ref{Primal Constraint 4})} needs to jointly determine the subset $\mathcal{A}$ and design the transmit covariance matrices $\mathbf{S}_{k,\mathcal{A}}$ for $K$ users, which is a combinatorial optimization problem and is NP-hard. A brute-force solution to a combinatorial optimization problem like (\ref{Primal Constraint 1})--(\ref{Primal Constraint 4}) is exhaustive search. Specifically, we must check all possible combinations of the active RAPs. For each combination, we must search for the optimal $\{\mathbf{S}_{k,\mathcal{A}}\}$ that satisfies the constraints (\ref{Primal Constraint 2})--(\ref{Primal Constraint 4}). In the end, we pick out the combination that maximizes the sum rate. However, the complexity grows exponentially with $L$, which cannot be applied to real-world application. Instead, we use the concept of $\ell_0$ norm to reformulate problem (\ref{Primal Constraint 1})--(\ref{Primal Constraint 4}). In particular, define $\bm{\omega}\in\mathbb{R}^{1\times L}$ as
\begin{equation}
\bm{\omega}=\left[\text{Tr}\{\mathbf{B}_1\sum_{k\in\mathcal{K}}\mathbf{S}_k\},\text{Tr}\{\mathbf{B}_2\sum_k\mathbf{S}_{k\in\mathcal{K}}\},\cdots,\text{Tr}\{\mathbf{B}_l\sum_{k\in\mathcal{K}}\mathbf{S}_k\},\cdots,\text{Tr}\{\mathbf{B}_L\sum_{k\in\mathcal{K}}\mathbf{S}_k\}\right],
\end{equation}
where $\mathbf{S}_k=\mathbb{E}\left[\mathbf{x}_k\mathbf{x}_k^\dag\right]$, with $\mathbf{x}_k\in\mathbb{C}^{M\times 1}$ denoting the transmit signal vector from all RAPs in $\mathcal{L}$ to user $k$. The $(N_c(l-1)+1)$-th to the $N_cl$-th entries of $\mathbf{x}_k$ are zero if RAP $l$ is inactive. It is clear that the $l$-th entry $\omega_l=\text{Tr}\{\mathbf{B}_l\sum_k\mathbf{S}_k\}$ is the transmit power of RAP $l$, which is non-zero if and only if RAP $l\in\mathcal{A}$. It~is easy to verify that $\|\bm{\omega}\|_0=|\mathcal{A}|$. We then have the following lemma:
\newtheorem{Th1}{Lemma}

\begin{Th1}\label{Lemma1}
{The problem defined in (\ref{Primal Constraint 1})--(\ref{Primal Constraint 4}) is} equivalent to the following optimization problem:%Please confirm.
\begin{equation}\label{P_l0}
\begin{aligned}
&\mathop{\text{maximize}}\limits_{\{\mathbf{S}_k\}}&&\sum_{k\in\mathcal{K}}\log_2\det\left(\mathbf{I}_N+\frac{1}{\sigma^2}\mathbf{H}_{k}\mathbf{S}_{k}\mathbf{H}_{k}^{\dag}\right) -\eta\|\bm\omega\|_0 &&\\
&\text{s.t.} && \mathbf{H}_j\mathbf{S}_k\mathbf{H}_j=\mathbf{0}&& \forall j,k\in\mathcal{K},j\ne k \\
&&& \text{Tr}\left\{\mathbf{B}_l\sum_{k\in\mathcal{K}}\mathbf{S}_{k}\right\}\leq P_{l,\max} && \forall l\in\mathcal{L} \\
&&& \mathbf{S}_{k}\succeq 0&& \forall k\in\mathcal{K}
\end{aligned}
\end{equation}
where $\mathbf{H}_k\in\mathbb{C}^{N\times M}$ is the channel gain matrix from all RAPs in $\mathcal{L}$ to user $k$, $k\in\mathcal{K}$.
\end{Th1}
\begin{proof}
Please refer to Appendix \ref{app_1} for detailed proof.
\end{proof}
Lemma \ref{Lemma1} indicates that instead of searching over the possible combinations of $\mathcal{A}$ and then optimizing according to the corresponding channel gain matrices $\{\mathbf{H}_{k,\mathcal{A}}\}$, (\ref{Primal Constraint 1})--(\ref{Primal Constraint 4}) can be solved based on the channel between the users and all RAP antennas, i.e., $\{\mathbf{H}_{k}\}$. However, the problem (\ref{P_l0}) is non-convex due to the existence of the $\ell_0$ norm, making it difficult to find the global optimal solution.

In compressive sensing theory, the $\ell_0$ norm is usually replaced by a $\ell_1$ norm, and sparse solution can be achieved. However, simply substituting $\|\bm\omega\|_0$ by $\|\bm\omega\|_1$ in (\ref{P_l0}) will not necessarily produce sparse solution in general, as $\|\bm\omega\|_1$ equals the sum power consumption instead of the number of non-zero entries. By replacing $\|\bm\omega\|_0$ by $\|\bm\omega\|_1$, the transmit power at all RAPs still tend to satisfy the power constraints with equality at the optimum, leading to a non-sparse solution. In this paper, we propose to solve (\ref{P_l0}) heuristically by iteratively relaxing the $\ell_0$ norm as a weighted $\ell_1$ norm. In particular, at the $t$-th iteration, {the $\ell_0$ norm $\|\bm\omega^{(t)}\|_0$ is approximated by}
%We replace the dot with multiplication sign in the equation, please confirm.
\begin{equation}\label{l0 approx}
\|\bm \omega^{(t)}\|_0\approx \sum_{l=1}^L \beta_l^{(t)}\omega_l^{(t)},
\end{equation}
where $\beta_l^{(t)}=\frac{1}{\text{Tr}\{\mathbf{B}_l\sum_{k}\mathbf{S}_k^{(t-1)}\}+\epsilon}$, with $\epsilon>0$ denoting a small positive constant. (\ref{P_l0}) can be then reformulated as
\begin{equation}\label{P_l0_approx}
\begin{aligned}
&\text{maximize}&& \sum_{k\in\mathcal{K}}\log_2\det\left(\mathbf{I}_N+\frac{1}{\sigma^2}\mathbf{H}_{k}\mathbf{S}_{k}\mathbf{H}_{k}^{\dag}\right)-\text{Tr}\left\{\bm{\Psi}^{(t)}\sum_{k\in\mathcal{K}}{\mathbf{S}_k}\right\} &&\\
&\text{s.t.} && \text{Constraints in (\ref{P_l0})},
\end{aligned}
\end{equation}
where
\begin{equation}\label{Psi}
\bm{\Psi}^{(t)}=\eta\sum_{l}\beta_l^{(t)}\mathbf{B}_l.
\end{equation}

{By noting that the first item of the objective function, i.e., $\sum_{k\in\mathcal{K}}\log_2\det\left(\mathbf{I}+\frac{1}{\sigma^2}\mathbf{H}_{k}\mathbf{S}_{k}\mathbf{H}_{k}^{\dag}\right)$, is~concave with respect to $\mathbf{S}_k$, and the second item $\text{Tr}\left\{\bm{\Psi}^{(t)}\sum_{k\in\mathcal{K}}{\mathbf{S}_k}\right\}$ is affine with respect to $\mathbf{S}_k$, we can then conclude that (\ref{P_l0_approx}) is a convex optimization problem. The problem } can be solved by standard convex optimization techniques, e.g. interior point method \cite{Boyd2004}, which, however, is typically slow. In fact, by utilizing the structure of BD precoding, the problem can be efficiently solved by its Lagrangian dual.

\section{Reweighted \boldmath{$\ell_1$} Based Algorithm}\label{s3}

In this section, the algorithm to solve the group sparse linear precoding problem will be presented. To solve (\ref{P_l0_approx}), it is desirable to remove the set of ZF constraints in the first place. It has been proved in~\cite{Zhang2010} that the optimal solution for BD precoding with per-RAP constraint is given by
\begin{equation}\label{Sopt_form}
\mathbf{S}_k=\mathbf{\tilde{V}}_k\mathbf{Q}_k\mathbf{\tilde{V}}_k^\dag,
\end{equation}
where $\mathbf{Q}_k\succeq 0$. $\mathbf{\tilde{V}}_k$ is given from the singular value decomposition (SVD) of $\mathbf{G}_k=[\mathbf{H}_1^T,\cdots,\mathbf{H}_{k-1}^T,\mathbf{H}_{k}^T,\cdots,\mathbf{H}_{K}^T]^T$ as
\begin{equation}
\mathbf{G}_k=\mathbf{U}_k\bm{\Sigma}_k[\mathbf{V}_k,\mathbf{\tilde{V}}_k]^\dag,
\end{equation}
where $\mathbf{\tilde{V}}_k\in\mathbb{C}^{M\times (M-N(K-1))}$ is the last $M-N(K-1)$ columns of the right singular matrix of $\mathbf{G}_k$. It is easy to verify that $\mathbf{H}_j\mathbf{S}_k\mathbf{H}_j^{\dag}=\mathbf{0}$ for all $j\ne k$.

Therefore, by substituting (\ref{Sopt_form}) into (\ref{P_l0_approx}), the problem reduces to
\begin{equation}\label{P_l0_approx1}
\begin{aligned}
&\text{maximize}&& \sum_{k\in\mathcal{K}}\log_2\det\left(\mathbf{I}_N+\frac{1}{\sigma^2}\mathbf{H}_{k}\mathbf{\tilde{V}}_k\mathbf{Q}_k\mathbf{\tilde{V}}_k^\dag\mathbf{H}_{k}^{\dag}\right)-\sum_{k\in\mathcal{K}}\text{Tr}\left\{\bm{\Psi}^{(t)}\mathbf{\tilde{V}}_k\mathbf{Q}_k\mathbf{\tilde{V}}_k^\dag\right\} &&\\
&\text{s.t.} && \sum_k\text{Tr}\left\{\mathbf{B}_l\mathbf{\tilde{V}}_k\mathbf{Q}_k\mathbf{\tilde{V}}_k^\dag\right\}\leq P_{l,\max} && \forall l\in\mathcal{L} \\
&&& \mathbf{Q}_{k}\succeq 0&& \forall k\in\mathcal{K}
\end{aligned}
\end{equation}

Note that (\ref{P_l0_approx1}) is a convex problem, its Lagrangian dual can be written as
\begin{align}
L\left(\{\mathbf{Q}_k\}_{k\in\mathcal{K}},\bm{\lambda}\right)&=\sum_{k\in\mathcal{K}}\log_2\det\left(\mathbf{I}_N+\frac{1}{\sigma^2}\mathbf{H}_{k}\mathbf{\tilde{V}}_k\mathbf{Q}_k\mathbf{\tilde{V}}_k^\dag\mathbf{H}_{k}^{\dag}\right) \nonumber \\
&-\sum_{k\in\mathcal{K}}\text{Tr}\left\{\bm{\Psi}^{(t)}\mathbf{\tilde{V}}_k\mathbf{Q}_k\mathbf{\tilde{V}}_k^\dag\right\}  \\
&-\sum_{l}\lambda_l\left(\sum_k\text{Tr}\left\{\mathbf{B}_l\mathbf{\tilde{V}}_k\mathbf{Q}_k\mathbf{\tilde{V}}_k^\dag\right\}-P_{l,\max}\right)\nonumber
\end{align}
where $\lambda_l\geq 0$ denoting the Lagrangian dual variables. The Lagrangian dual function can be given as
\begin{equation}
g(\bm{\lambda})=\max_{\mathbf{Q}_k\succeq 0, k\in\mathcal{K}} L\left(\{\mathbf{Q}_k\}_{k\in\mathcal{K}},\bm{\lambda}\right),
\end{equation}
where $\bm\lambda=[\lambda_1,\lambda_2,\cdots,\lambda_L]$. We can then obtain the Lagrangian dual problem of (\ref{P_l0_approx1}) as
\begin{equation}\label{P_dual}
\min_{\bm\lambda\geq\mathbf{0}} g(\bm{\lambda}).
\end{equation}

Since the problem (\ref{P_l0_approx1}) is convex and satisfies the Slater's condition, strong duality holds. The~respective primal and dual objective values in (\ref{P_l0_approx1}) and (\ref{P_dual}) must be equal at the global optimum, and the complementary slackness must hold at the optimum, i.e.,
\begin{equation}\label{com_slack}
\lambda_l^*\left(\sum_{k\in\mathcal{K}}\text{Tr}\left\{\mathbf{B}_l\mathbf{\tilde{V}}_k\mathbf{Q}^*_k\mathbf{\tilde{V}}_k^\dag\right\}-P_{l,\max}\right)=0,\;l=1,\cdots,L,
\end{equation}
where $\{\mathbf{Q}_k^*\}$ and  $\{\lambda_l^*\}$ are the optimal primal and dual variables, respectively.

For fixed $\bm\lambda$, the Lagrangian dual function $g(\bm{\lambda})$ can be obtained by solving
\begin{equation} \label{fix_dual}
\max_{\mathbf{Q}_k\succeq 0} \sum_{k\in\mathcal{K}}\log_2\det\left(\mathbf{I}_N+\frac{1}{\sigma^2}\mathbf{H}_{k}\mathbf{\tilde{V}}_k\mathbf{Q}_k\mathbf{\tilde{V}}_k^\dag\mathbf{H}_{k}^{\dag}\right)-\sum_k\text{Tr}\left\{\bm{\Omega}\mathbf{\tilde{V}}_k\mathbf{Q}_k\mathbf{\tilde{V}}_k^\dag\right\} ,
\end{equation}
where $\bm{\Omega}=\bm{\Psi}^{(t)}+\sum_{l}{\lambda_l\mathbf{B}_l}$. {Appendix \ref{app_2} shows that the optimal $\mathbf{{Q}}_k^*$ can be obtained as
\begin{equation}\label{Q_opt}
\mathbf{{Q}}_k^*=\left(\mathbf{\tilde{V}}_k^\dag\bm{\Omega}\mathbf{\tilde{V}}_k\right)^{-1/2}\mathbf{\hat{V}}_k^{\dag}\bm{\tilde{\Lambda}}_k\mathbf{\hat{V}}_k\left(\mathbf{\tilde{V}}_k^\dag\bm{\Omega}\mathbf{\tilde{V}}_k\right)^{-1/2},
\end{equation}
where $\mathbf{\hat{V}}_k$ is obtained from the following reduced SVD:
\begin{equation}\label{svd1}
\mathbf{H}_{k}\mathbf{\tilde{V}}_k\left(\mathbf{\tilde{V}}_k^\dag\bm{\Omega}\mathbf{\tilde{V}}_k\right)^{-1/2}=\mathbf{\hat{U}}_k\bm{\Xi}_k\mathbf{\hat{V}}_k^\dag.
\end{equation}
$\bm{\tilde{\Lambda}}_k=\text{diag}\{\tilde{\lambda}_{k,1},\cdots,\tilde{\lambda}_{k,N}\}$, with
\begin{equation}
\tilde{\lambda}_{k,n}=\left(\frac{1}{\ln 2}-\frac{\sigma^2}{\xi_{k,n}^2}\right)^+,
\end{equation}
where $x^+=\max (x,0)$.}

The optimal $\mathbf{S}_k^*$ for given $\bm{\lambda}$ can be then obtained as
\begin{equation}\label{Covariance}
\mathbf{S}_k^*(\bm\lambda)=\mathbf{\tilde{V}}_k\left(\mathbf{\tilde{V}}_k^\dag\bm{\Omega}\mathbf{\tilde{V}}_k\right)^{-1/2}\mathbf{\hat{V}}_k^{\dag}\bm{\tilde{\Lambda}}_k\mathbf{\hat{V}}_k\left(\mathbf{\tilde{V}}_k^\dag\bm{\Omega}\mathbf{\tilde{V}}_k\right)^{-1/2}\mathbf{\tilde{V}}_k^\dag
\end{equation}

With the optimal $\mathbf{S}_k^*(\bm\lambda)$ achieved for given $\bm{\lambda}$, we can then find the Lagrangian dual variables $\bm\lambda$ by the projected subgradient method. Projected subgradient methods following, e.g., the square summable but not summable step size rules, have been proven to converge to the optimal values \cite{Bertsekas2003}. In particular, a subgradient with respect to $\lambda_l$ is $P_{l,\max}-\sum_{k\in\mathcal{K}}\text{Tr}\left\{\mathbf{B}_l\mathbf{\tilde{V}}_k\mathbf{Q}_k\mathbf{\tilde{V}}_k^\dag\right\}$. With a step size $\delta_t$, the~dual variables can be updated as
\begin{equation}\label{subgradient}
\lambda_l^{(t+1)}=\max\left\{\lambda_l^{(t)}-\delta_t \left(P_{l,\max}-\sum_{k\in\mathcal{K}}\text{Tr}\left\{\mathbf{B}_l\mathbf{\tilde{V}}_k\mathbf{Q}_k\mathbf{\tilde{V}}_k^\dag\right\}\right),0\right\},\;\forall l=1,\cdots,L.
\end{equation}

From the complementary slackness in (\ref{com_slack}), a stopping criterion for updating (\ref{subgradient}) can be
\begin{equation}\label{com_slack_sum}
\sum_{l\in\mathcal{L}}|\lambda_l\sum_{k\in\mathcal{K}}\text{Tr}\left\{\mathbf{B}_l\mathbf{\tilde{V}}_k\mathbf{Q}_k\mathbf{\tilde{V}}_k^\dag\right\}|^2<\varepsilon.
\end{equation}

Once the optimal $\bm{\lambda}^*$ is obtained, the optimal precoding matrices $\{\mathbf{T}_k^*\}$ can be achieved by using the fact that $\mathbf{S}_k=\mathbf{T}_k\mathbf{T}_k^{\dag}$ as
\begin{equation}
\mathbf{T}_k^*=\mathbf{\tilde{V}}_k\left(\mathbf{\tilde{V}}_k^\dag\bm{\Omega}^*\mathbf{\tilde{V}}_k\right)^{-1/2}\mathbf{\hat{V}}_k^{\dag}\bm{\Lambda}_k^{1/2}.
\end{equation}

The optimal set $\mathcal{A}$ and the corresponding precoding matrices $\{\mathbf{T}_{k,\mathcal{A}}\}$ can be obtained from $\{\mathbf{T}_k^*\}$. The~algorithm is summarized as Algorithm \ref{Al1}.

\begin{algorithm}[H]
\caption{Reweighted $\ell_1$ Norm Based Sparse Precoding Design.}\label{Al1}
\textbf{Initilization:} Set iteration counter $t=1$, Lagrangian dual variable $\bm{\lambda}^{(0)}>0$, $l=1,\cdots,L$. \\
\textbf{Repeat:}
\begin{enumerate}
  \item Calculate $\bm{\Psi}^{(t)}$ and $\mathbf{S}_k^{(t)}$, $k=1,\cdots, K$, according to (\ref{Psi}) and (\ref{Covariance}), respectively.
  \item Update $\{\lambda_l\}$ according to (\ref{subgradient}).
  \item Update the iteration counter {$t\leftarrow t+1$} and %Please check it.
  \[
  r=\sum_{l\in\mathcal{L}}|\lambda_l\sum_{k\in\mathcal{K}}\text{Tr}\left\{\mathbf{B}_l\mathbf{\tilde{V}}_k\mathbf{Q}_k\mathbf{\tilde{V}}_k^\dag\right\}|^2.
  \]
\end{enumerate}
\textbf{Stop} if $r<\varepsilon$, where $\varepsilon$ is a pre-defined tolerance threshold.
\label{alg_1}
\end{algorithm}

\section{Complexity Analysis}\label{s4}
{In this section, we provide our complexity analysis of Algorithm \ref{alg_1}. Note that $\mathbf{\tilde{V}}_k$ can be calculated before the iterations. The main complexity of the proposed algorithm lies in step 1 and step 2.}

{Let us first focus on step 1. According to \cite{Coppersmith1987}, the calculations of matrix multiplication $\mathbf{XY}$ and matrix inversion $\mathbf{Z}^{-1}$ have complexities on the orders of $O(mnp)$ and $O(m^{2.736})$, respectively, where~$\mathbf{X}\in\mathbb{C}^{m\times n}$, $\mathbf{Y}\in\mathbb{C}^{n\times p}$ and $\mathbf{Z}\in\mathbb{C}^{m\times m}$. Therefore, the complexity to calculate $\mathbf{H}_{k}\mathbf{\tilde{V}}_k\left(\mathbf{\tilde{V}}_k^\dag\bm{\Omega}\mathbf{\tilde{V}}_k\right)^{-1/2}$ is on the order of $O\left((M-N(K-1))^{2.736}\right)$. On the other hand, as the complexity of the SVD in (\ref{svd1}) is on the order of $O\left(M(M-N(K-1))^2\right)$ \cite{Coppersmith1987}, the complexity to calculate $\mathbf{\hat{V}}_k$ is on the order of $O\left(M(M-N(K-1))^2\right)$. By considering that the matrix calculation in (\ref{Covariance}) has a complexity on the order of $O\left(M(M-N(K-1))^2\right)$, we can then conclude that the complexity of step 1 is on the order of $O\left(KM(M-N(K-1))^2\right)$.}

{In step 2, as $\sum_{k\in\mathcal{K}}\text{Tr}\left\{\mathbf{B}_l\mathbf{\tilde{V}}_k\mathbf{Q}_k\mathbf{\tilde{V}}_k^\dag\right\}=\text{Tr}\left\{\mathbf{B}_l\sum_{k\in\mathcal{K}}\mathbf{S}_k\right\}$ and $\sum_{k\in\mathcal{K}}\mathbf{S}_k$ can be calculated before updating $\lambda_l$, step 2 has a complexity on the order of $O\left(LM^3\right)$. By noting that the number of RAPs is typically larger than the number of users, i.e., $L>K$, and that the total number of BS antennas $M=N_c L$, the overall complexity of the proposed algorithm is on the order of $O(t_{avg}N_c^3L^4)$, where $t_{avg}$ is the average number of iterations. As we will show in the following section, the algorithm is able to converge within 15 iterations for a properly selected step size and tolerance threshold.}

\section{Simulation Results}\label{s5}

In this section, simulation results are presented to illustrate the results in this paper. We assume that the channel gain matrices $\{\mathbf{H}_{k,l}\}$ from RAP $l$ to user $k$ are independent over $k$ and $l$ for all $l\in\mathcal{L}$ and $k\in\mathcal{K}$, and all entries of $\mathbf{H}_{k,l}$ are independent and identically distributed (iid) complex Gaussian random variables with zero mean and variance $\gamma_{k,l}^2$. The path-loss model from RAP $l$ to the user $k$ is
\begin{equation}
\text{PL}_{k,l}\text{(dB)}=128+37.6\log_{10}D_{k,l},
\end{equation}
$\forall l\in\mathcal{L}$ and $\forall k\in\mathcal{K}$, where $D_{k,l}$ is the distance between user $k$ and RAP $l$ in the unit of kilometer. The~large-scale fading coefficient from RAP $l$ to user $k$ can be then obtained as
\begin{equation}
\gamma_{k,l}^2=10^{-\text{PL}_{k,l}\text{(dB)}/10}.
\end{equation}

The transmit power constraint at each RAP are assumed to be identical, which is set to be $P_{l,\max}=-40$ dBm/Hz, $\forall l\in\mathcal{L}$, and the noise variance is set to be $\sigma^2=-162$ dBm/Hz.

\begin{figure}[H]
\begin{center}
\includegraphics[width=.65\textwidth]{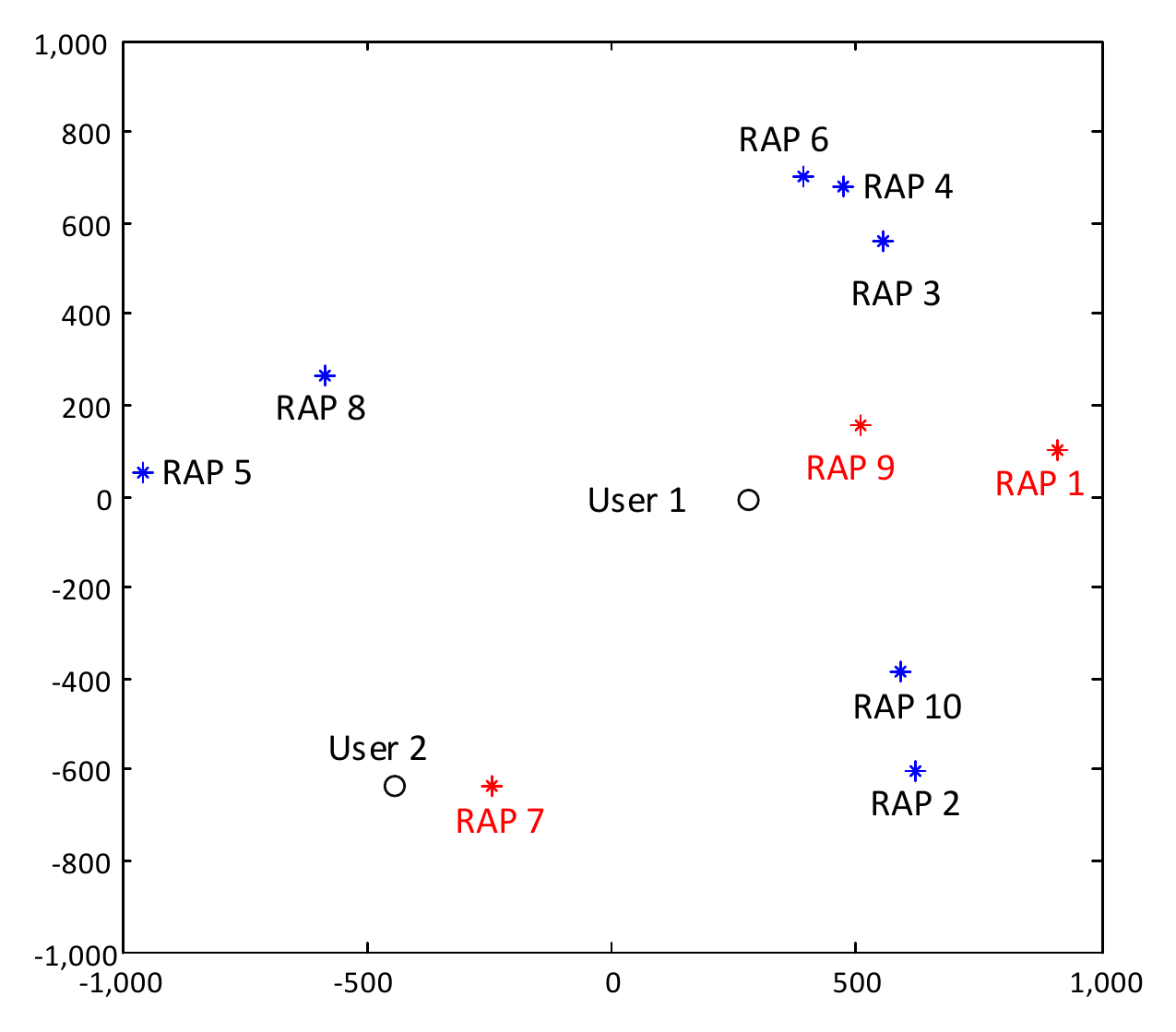}
\vspace{-1\baselineskip}
\caption{{An example of a randomly generated} system configuration. The active RAPs at the converged state with $\eta=0.5$ are highlighted in red. }\label{Fig_layout}
%Please add separators to the numbers sucn as (1,000 and -1,000) in the figugre, and make sure the figure is high-res figure (at least 300 dpi).
\end{center}
\end{figure}

\begin{figure}[H]
\begin{center}
\includegraphics[width=.6\textwidth]{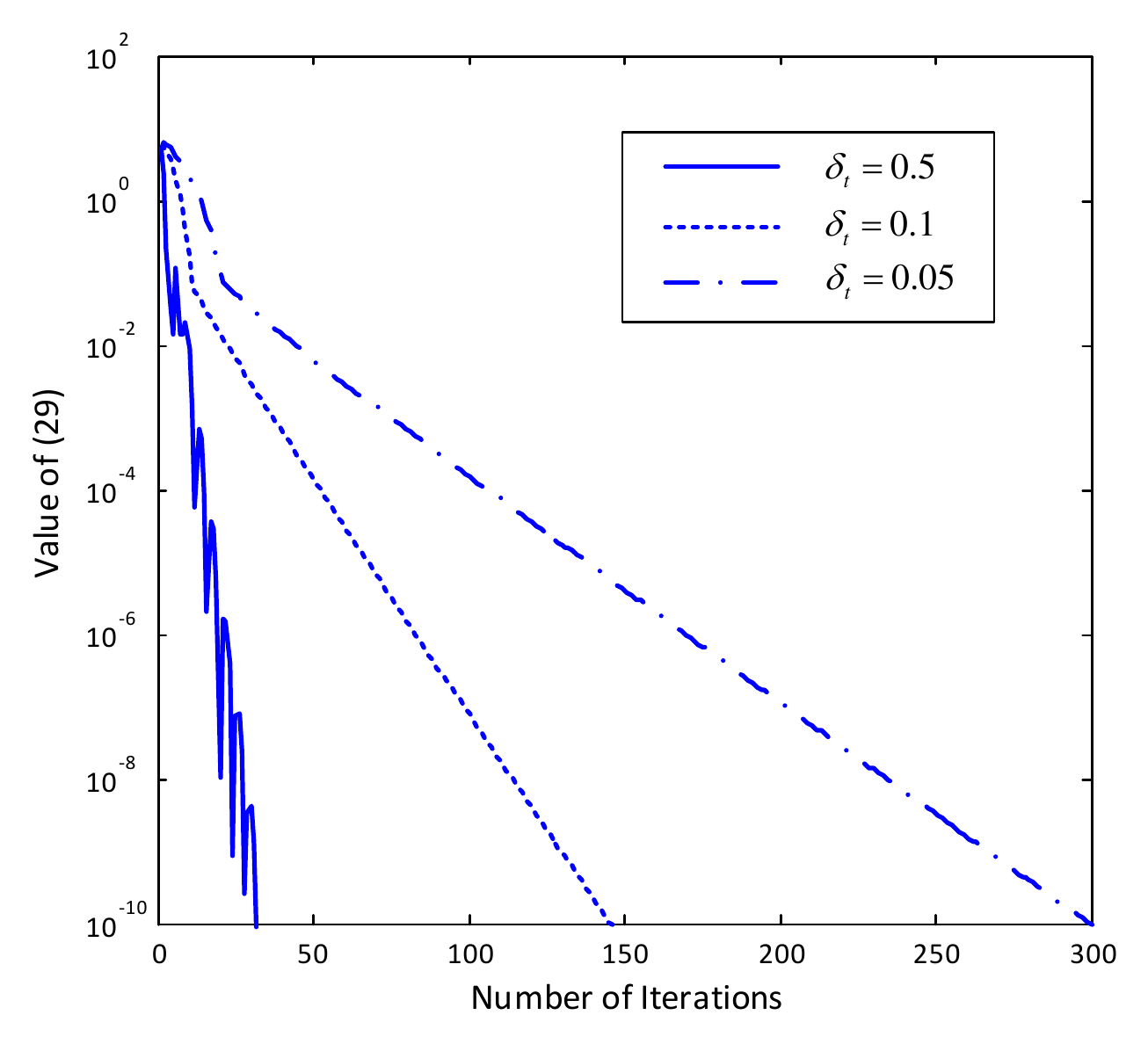}
\vspace{-1\baselineskip}
\caption{{Convergence behavior of the proposed algorithm} under the antenna and user layout shown in Figure~\ref{Fig_layout} with randomly generated small-scale fading coefficients.  $L=10$. $N_c=2$. $K=2$. $N=3$. $\eta=0.5$. }\label{Fig_converge}
\end{center}
\end{figure}

We consider the case that $L=10$ RAPs with $N_c=2$ antennas each and $K=2$ users with $N=3$ antennas each. {The positions of users and RAPs are generated in a circular area with radius 1,000~km %Please confirm the separator ``,''
 following uniform distribution. An example of the randomly generated antenna and user layout is plotted in Figure~\ref{Fig_layout}. Figure~\ref{Fig_converge} shows the convergence behavior of the proposed algorithm under the layout given in Figure~\ref{Fig_layout} under different step size $\delta_t=0.5$, $0.1$ and $0.05$. As we can see from Figure~\ref{Fig_converge}, the~value of (\ref{com_slack_sum}) rapidly decreases with the number of iterations. When the step size is large, i.e., $\delta_t=0.5$, despite that the algorithm converges in general, the value increases after several iterations. Such observation comes from the fact that we use the reweighted $\ell_1$ norm to approximate the non-convex $\ell_0$ norm, and a large step in $\bm\lambda$ will generally leading to a significant change of $\bm\Psi$ in~(\ref{P_l0_approx1}). When the step size is smaller, $\bm\Psi$ changes moderately, and a monotonic decrease can be observed in Figure~\ref{Fig_converge} when $\delta_t=0.1$ and $0.05$.}

\begin{figure}[H]
\begin{center}
\includegraphics[width=.6\textwidth]{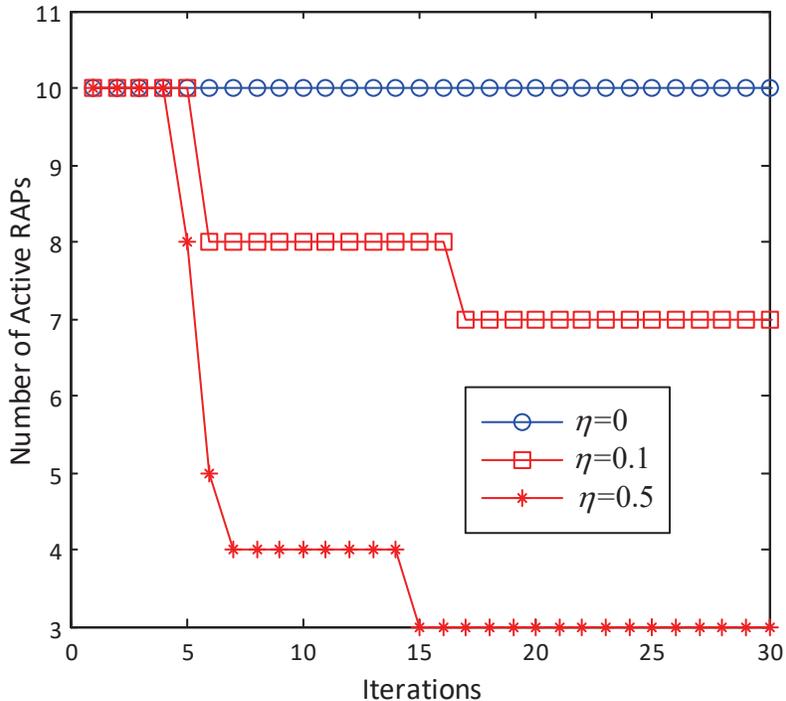}
\vspace{-1\baselineskip}
\caption{Number of active RAPs under the antenna and user layout shown in Figure~\ref{Fig_layout} with randomly generated small-scale fading coefficients. $L=10$. $N_c=2$. $K=2$. $N=3$. }\label{Fig_active}
\end{center}
\end{figure}

{Figure~\ref{Fig_active} shows how the number of active RAPs varies with iterations in the layout shown in Figure~\ref{Fig_converge} with step size $\delta_t=0.1$ and the tolerance threshold $\varepsilon=10^{-4}$, {where a RAP is said to be active if its transmit power $P_l\geq 10^{-5}P_{l,\max}$}. The tradeoff constant $\eta$ is set to be $0$, $0.1$ and $0.5$.} Figure~\ref{Fig_active} shows that the number of active RAPs $|\mathcal{A}|$ decreases with $\eta$. Specifically, with~$\eta=0$, the problem reduces to a BD precoder design with per-RAP constraint, and all RAPs are active. With $\eta=0.5$, the sparsest solution can be achieved, i.e., $|\mathcal{A}|=A_{\min}=\lceil\frac{KN}{N_c}\rceil=3$. We~should mention that the value of $\eta$ that corresponds to the sparsest solution varies with the system~configuration. % footnote moved to the main text according to the jounal style, please confirm

\begin{figure}[H]
\begin{center}
\includegraphics[width=.7\textwidth]{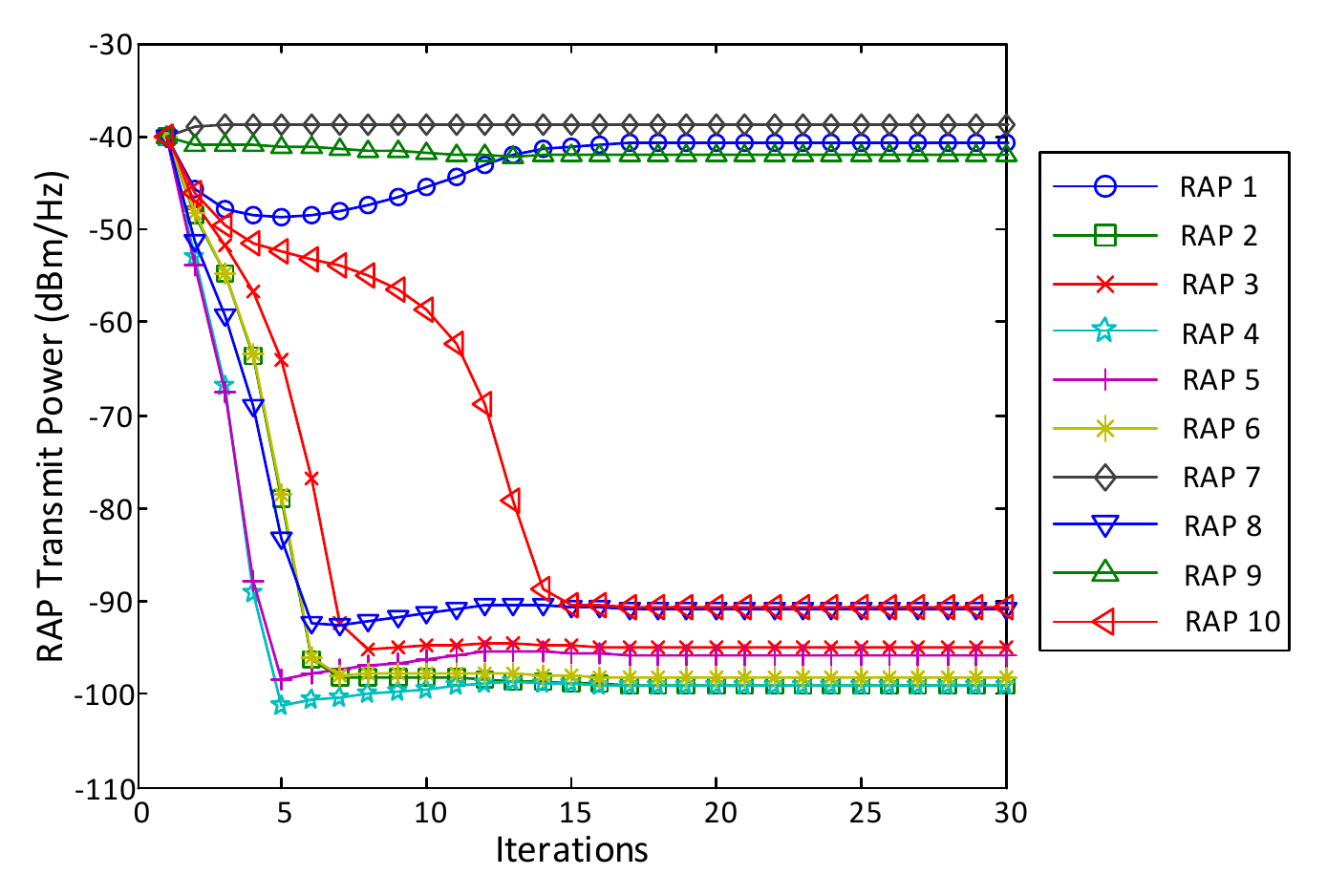}
\vspace{-1\baselineskip}
\caption{{Transmit power of the RAPs under the antenna} and user layout shown in Figure~\ref{Fig_layout} with randomly generated small-scale fading coefficients.  $L=10$. $N_c=2$. $K=2$. $N=3$. $\eta=0.5$.}\label{Fig_bs}
%please change the hyphen in the figure into minus sign.
\end{center}
\end{figure}

As Figure~\ref{Fig_active} shows, the first several iterations lead to the biggest improvement. As iterations go on, there is no further improvement after the 15th iteration. Compared to the full cooperation case, i.e., $\eta=0$, as $L-A_{\min}=7$ RAPs are switched off, 70\% of the circuit power consumption can be saved with, however, limited rate performance loss, which will be illustrated later in this section. Figure~\ref{Fig_bs}~plots the transmit power distribution over all 10 RAPs. {Intuitively, the RAPs that have the smaller access distances contribute most to the sum rate. As shown in Figures~\ref{Fig_layout} and \ref{Fig_bs}, RAP 1 and RAP 9, which are close to User 1, and RAP 7, which is close to User 2, are selected to be active, while all the other RAP's transmit power eventrally goes to zero after 15 iterations.}

\begin{figure}[H]
\begin{center}
\includegraphics[width=.6\textwidth]{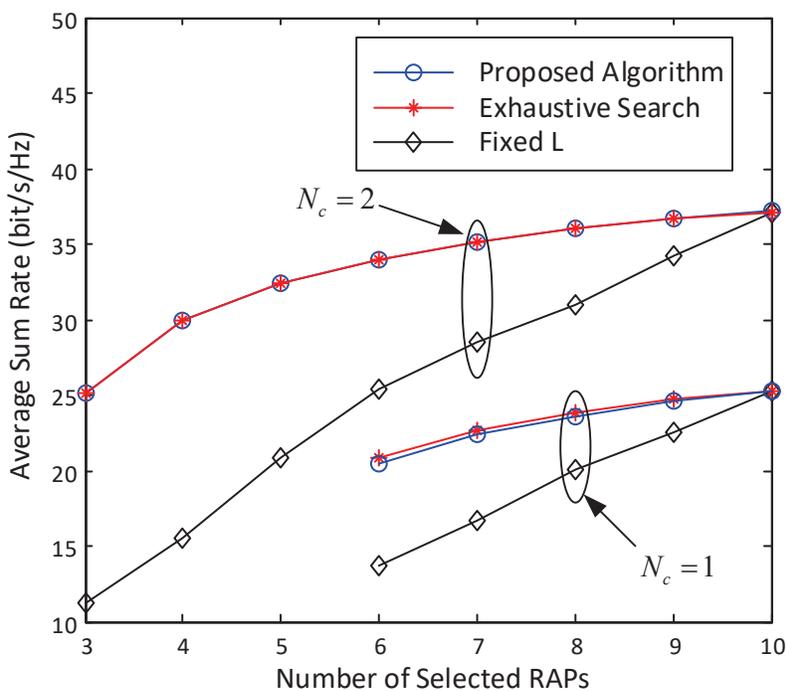}
\vspace{-1\baselineskip}
\caption{{Tradeoff between average sum rate and the number} of active RAPs. $L=10$. $K=2$. $N=3$.}\label{Fig_tradeoff}
\end{center}
\end{figure}

Figure~\ref{Fig_tradeoff} plots how the average sum rate varies with the number of active RAPs. The average sum rate is obtained by averaging over 20 realizations of small-scale fading and 30 realizations of the positions of RAPs and users. The results of the exhaustive search is also presented for comparison, which~is obtained by searching over all possible combinations of active RAPs, and computing its achievable sum rate by the algorithm given in \cite{Zhang2010}. For the proposed algorithm, we simulate a series of different $\eta$'s to get different points along the curve. As we can see from Figure~\ref{Fig_tradeoff}, our proposed algorithm achieves almost the same average sum rate as the exhaustive search, which verifies the optimality of our proposed algorithm. Figure~\ref{Fig_tradeoff} further plots the average sum rate with a fixed number of $|\mathcal{A}|$ instead of $L$ uniformly distributed antennas are deployed for comparison, which is denoted as ``Fixed $L$'' in Figure~\ref{Fig_tradeoff}. We can clearly see that the proposed algorithm can achieve much better rate performance over that with the same amount of transmit RAP. With six selected antennas, for instance, the group sparse precoder reduces the average sum rate for only 3 bit/s/Hz, whereas if only $6$ antennas were installed instead of $L=10$, an additional 9 bit/s/Hz rate loss can be observed compared to the proposed algorithm. Moreover, the rate gap between the proposed algorithm and that with a fixed number of $L=|\mathcal{A}|$ full cooperative RAPs further increases as the number of selected antennas $|\mathcal{A}|$ decreases. This highlights the importance of group sparse precoder design in C-RAN with a large number of distributed RAPs.

\section{Conclusions}\label{s6}

In this paper, we study the group sparse precoder design that maximizes the sum rate in a C-RAN. We show that the joint antenna selection and precoder design problem can be formulated into an $\ell_0$ norm problem, which is, however, combinatorial and NP-hard. Inspired by the theory of compressive sensing, we propose an approach that solves the problem via reweighted $\ell_1$ norm. Simulation results verify the optimality of our proposed algorithm in that it achieves almost the same performance as that obtained from the exhaustive search. Compared to full cooperation, the group sparse precoding can achieve a significant proportion of the maximum sum rate that was achieved from full cooperation with, however, much fewer active RAPs, which highlights the importance of employing group sparse precoding in C-RAN with ultra-dense RAPs.

Note that in practical system, the imperfect fronthaul links between the BBU pool and the RAPs further limits the performance. It is desirable to extend our approach to a practical scenario by including the fronthaul link capacity constraint. Moreover, inspired by \cite{Vu2016} which implemented the zero-forcing precoding {in a distributed manner, we would like to further extend our algorithm in a distributed way in our future work, as it enables the application of group sparse precoding in a large-scale C-RAN system.}

\appendices
\section{Proof of Lemma \ref{Lemma1}}\label{app_1}
Define the channel gain matrix and the precoding matrix between user $k$ and the antennas of RAP $l$ as $\mathbf{H}_{k,l}$ and $\mathbf{T}_{k,l}$, respectively, $\forall k\in\mathcal{K}$ and $\forall l\in\mathcal{L}$. Let us reorder the channel gain matrices $\mathbf{H}_k$ and the precoding matrices $\mathbf{T}_k$ as
\begin{equation}
\mathbf{\tilde{H}}_k=\left[[\mathbf{H}_{k,l}]_{l\in\mathcal{A}},[\mathbf{H}_{k,l}]_{l\not\in\mathcal{A}}\right],
\end{equation}
and
\begin{equation}
\mathbf{\tilde{T}}_k=\left[[\mathbf{T}_{k,l}^T]_{l\in\mathcal{A}},[\mathbf{T}_{k,l}^T]_{l\not\in\mathcal{A}}\right]^T,
\end{equation}
respectively. We have
\begin{align}\label{app hsh}
\mathbf{H}_j\mathbf{S}_k\mathbf{H}_j&=\mathbf{\tilde{H}}_k\mathbf{\tilde{T}}_k\mathbf{\tilde{T}}_k^\dag\mathbf{\tilde{H}}_k^\dag\nonumber \\
&\mathop{=}\limits^{(a)}\left(\sum_{l\in\mathcal{A}}\mathbf{H}_{k,l}\mathbf{T}_{k,l}\right)\left(\sum_{l\in\mathcal{A}}\mathbf{H}_{k,l}\mathbf{T}_{k,l}\right)^\dag  \\
&=\mathbf{H}_{j,\mathcal{A}}\mathbf{S}_{k,\mathcal{A}}\mathbf{H}_{j,\mathcal{A}}^\dag, \nonumber
\end{align}
for all $j,k\in\mathcal{K}$, where (a) follows from the fact that $\mathbf{T}_{k,l}=\mathbf{0}$ if RAP $l$ is not active, i.e., $l\not\in \mathcal{A}$.

Let us then prove the positive semidefinite constraint. The reordered covariance matrix $\mathbf{S}_k$ can be written as
\begin{align}\label{app psd}
\mathbf{\hat{S}}_k&=\mathbf{\hat{T}}_k\mathbf{\hat{T}}_k^{\dag} \nonumber \\
&=\begin{bmatrix}
    \mathbf{T}_{k,\mathcal{A}}\mathbf{T}_{k,\mathcal{A}}^\dag & \mathbf{T}_{k,\mathcal{A}}[\mathbf{T}_{k,l}]^\dag_{l\not\in\mathcal{A}} \\
    [\mathbf{T}_{k,l}]_{l\not\in\mathcal{A}}\mathbf{T}_{k,\mathcal{A}}^\dag & [\mathbf{T}_{k,l}]_{l\not\in\mathcal{A}}[\mathbf{T}_{k,l}]^\dag_{l\not\in\mathcal{A}}\\
  \end{bmatrix}
   \\
  &\mathop{=}\limits^{(b)}\begin{bmatrix}
      \mathbf{T}_{k,\mathcal{A}}\mathbf{T}_{k,\mathcal{A}}^\dag & \mathbf{0} \\
      \mathbf{0} & \mathbf{0} \\
    \end{bmatrix},\nonumber
\end{align}
where (b) holds as $\mathbf{T}_{k,l}=\mathbf{0}$ for all $l\not\in\mathcal{A}$. By noting that $\mathbf{T}_{k,\mathcal{A}}\mathbf{T}_{k,\mathcal{A}}^\dag=\mathbf{S}_{k,\mathcal{A}}\succeq \mathbf{0}$, we immediately have $\mathbf{S}_k\succeq 0$. (\ref{P_l0}) can be then proved by combining (\ref{app hsh})--(\ref{app psd}) and (\ref{Primal Constraint 1})--(\ref{Primal Constraint 4}).

\section{Derivation of (\ref{Q_opt})}\label{app_2}
{
We can observe from (\ref{fix_dual}) that the problem is equivalent to solve $K$ uncoupled subproblems given~as
\begin{equation}\label{app_subproblem}
  \max_{\mathbf{Q}_k\succeq 0} \log_2\det\left(\mathbf{I}_N+\frac{1}{\sigma^2}\mathbf{H}_{k}\mathbf{\tilde{V}}_k\mathbf{Q}_k\mathbf{\tilde{V}}_k^\dag\mathbf{H}_{k}^{\dag}\right)-\text{Tr}\left\{\bm{\Omega}\mathbf{\tilde{V}}_k\mathbf{Q}_k\mathbf{\tilde{V}}_k^\dag\right\} ,
\end{equation}

By noting that $\text{Tr}\{\mathbf{X}\mathbf{Y}\}=\text{Tr}\{\mathbf{Y}\mathbf{X}\}$, we have
\begin{equation}
\text{Tr}\left\{\bm{\Omega}\mathbf{\tilde{V}}_k\mathbf{Q}_k\mathbf{\tilde{V}}_k^\dag\right\}=\text{Tr}\left\{\left(\mathbf{\tilde{V}}_k^\dag\bm{\Omega}\mathbf{\tilde{V}}_k\right)^{1/2}\mathbf{Q}_k\left(\mathbf{\tilde{V}}_k^\dag\bm{\Omega}\mathbf{\tilde{V}}_k\right)^{1/2}\right\}.
\end{equation}

Let us define
\begin{equation}\label{app_qtilde_def}
  \mathbf{\tilde{Q}}_k=\left(\mathbf{\tilde{V}}_k^\dag\bm{\Omega}\mathbf{\tilde{V}}_k\right)^{1/2}\mathbf{Q}_k\left(\mathbf{\tilde{V}}_k^\dag\bm{\Omega}\mathbf{\tilde{V}}_k\right)^{1/2}.
\end{equation}

(\ref{app_subproblem}) can be written as
\begin{equation}\label{fix_dual 1}
\max_{\mathbf{\tilde{Q}}_k\succeq 0} \log_2\det\left(\mathbf{I}_N+\frac{1}{\sigma^2}\mathbf{H}_{k}\mathbf{\tilde{V}}_k\left(\mathbf{\tilde{V}}_k^\dag\bm{\Omega}\mathbf{\tilde{V}}_k\right)^{-1/2}\mathbf{\tilde{Q}}_k\left(\mathbf{\tilde{V}}_k^\dag\bm{\Omega}\mathbf{\tilde{V}}_k\right)^{-1/2}\mathbf{\tilde{V}}_k^\dag\mathbf{H}_{k}^{\dag}\right)
-\text{Tr}\left\{\mathbf{\tilde{Q}}_k\right\}.
\end{equation}

(\ref{fix_dual 1}) can be solved from standard waterfilling algorithm \cite{Cover2006}. In particular, by introducing the (reduced) SVD:
\begin{equation}
\mathbf{H}_{k}\mathbf{\tilde{V}}_k\left(\mathbf{\tilde{V}}_k^\dag\bm{\Omega}\mathbf{\tilde{V}}_k\right)^{-1/2}=\mathbf{\hat{U}}_k\bm{\Xi}_k\mathbf{\hat{V}}_k^\dag,
\end{equation}
where $\mathbf{\hat{U}}_k\in\mathbb{C}^{N\times N}$ and $\mathbf{\hat{V}}_k\in\mathbb{C}^{N\times N}$ are unitary matrices. $\bm{\Xi}_k=\text{diag}\{\xi_{k,1},\dots,\xi_{k,N}\}$, with $\xi_{k,n}$ denoting the $n$-th singular value of $\mathbf{H}_{k}\mathbf{\tilde{V}}_k\left(\mathbf{\tilde{V}}_k^\dag\bm{\Omega}\mathbf{\tilde{V}}_k\right)^{-1/2}$.
The optimal $\mathbf{\tilde{Q}}_k$ can be then obtained as:
\begin{equation}\label{Qtilde}
\mathbf{\tilde{Q}}_k^*=\mathbf{\hat{V}}_k^{\dag}\bm{\tilde{\Lambda}}_k\mathbf{\hat{V}}_k,
\end{equation}
where $\bm{\tilde{\Lambda}}_k=\text{diag}\{\tilde{\lambda}_{k,1},\cdots,\tilde{\lambda}_{k,N}\}$, with
\begin{equation}
\tilde{\lambda}_{k,n}=\left(\frac{1}{\ln 2}-\frac{\sigma^2}{\xi_{k,n}^2}\right)^+,
\end{equation}
where $x^+=\max(x,0)$.
}

{(\ref{Q_opt}) can be then obtained by combining (\ref{Qtilde}) and (\ref{app_qtilde_def}).}

\bibliography{ref}

\end{document}